# Möbius-like Real-Space Topology Reshapes Spectral Winding Topology in Hatano–Nelson Rings

*Yekai Shen[1,2,†], Shuhang Chen[2,†], Zishun Liao[2], Zhipeng Li[2,*]*

[1]School of the Gifted Young, University of Science and Technology of China, Hefei, China

[2] School of Information Science and Technology, University of Science and Technology of China, Hefei, China

*Email address for corresponding author: zhipeng.li@ustc.edu.cn;

[†]These authors contributed equally to this work.

***Abstract*—The spectral winding number serves as a bulk topological invariant in non-Hermitian systems, governing the emergence of skin modes and encoding the non-Hermitian bulk–boundary correspondence. However, most existing studies are built on conventional lattice geometries such as linear chains, rings, or planar arrays, leaving the role of real-space topological connectivity as an independent degree of freedom largely unexplored. Here, we construct a Möbius ring system by cutting two parallel Hatano–Nelson (HN) rings and reconnecting them with a half-twist, without altering any local hopping parameter. This topological reconstruction transforms the periodic-boundary spectrum from two disjoint ellipses into a multi-petalled rose curve, and leads to distinct decay lengths for different eigenstates under open boundary conditions. Moreover, the spectral winding number can be driven through discrete winding-number jumps by tuning the coupling strength, with critical values obtained analytically. Our results demonstrate that real-space Möbius connectivity, mediated by the coupling strength, provides an independent and tunable foundation for the systematic control of non-Hermitian topology, with implications for the design of topological devices and sensing schemes.**

## I. INTRODUCTION

Non-Hermitian Hamiltonians provide a natural and efficient description for open and dissipative systems such as those with gain, loss, or non-reciprocity, and have attracted intense theoretical and experimental interest across quantum optics, photonics, acoustics, ultracold atoms, and circuit metamaterials [1–6]. In non-Hermitian systems, the Bloch spectrum under periodic boundary conditions (PBCs) forms closed loops in the complex energy plane, and the winding of such a loop around a reference energy defines an integer point-gap topological invariant, the spectral winding number [7–11]. Serving as the bulk topological invariant in the non-Hermitian bulk–boundary correspondence, this winding number has been rigorously shown to determine both the number and the localization direction of skin modes under open boundary conditions, constituting a directly quantized physical observable [10]. The nontrivial point-gap topology manifests under open boundary conditions as a macroscopic accumulation of bulk eigenstates at the system boundary, the non-Hermitian skin effect (NHSE) [12–15]. Experimental observations of the NHSE across a wide range of classical and quantum platforms [16–22] have further inspired proposals for applications such as topological sensing and nonreciprocal wave funneling [23,24].

Engineering spectral winding in non-Hermitian chains has so far relied mainly on local parameters that carry or sustain directional bias in the bulk, such as tuning local or long-range non-reciprocal hopping strengths [25–28]. By comparison, how lattice sites are topologically connected in real space, as an alternative organizing principle for spectral winding, has received relatively little attention. Möbius geometries furnish a prominent instance of such a construction. In photonics and acoustics, Möbius-strip-inspired structures have been shown to host topological phases with globally non-trivial character, from projected Möbius edge bands to geometric-phase effects in resonant cavities [29–33]. Whether such a real-space topological reconnection can directly reorganize the spectral winding topology in a predictable and quantized manner has remained unaddressed.

To address this, we start from two parallel HN rings, cut each at a single bond, and reconnect the four free ends with a half-twist, obtaining a Möbius-reconnected ring in which the two sub-chains merge into a single closed loop with a consistent non-reciprocal flow, as illustrated in Fig. 1. All hopping amplitudes and on-site parameters are identical between the two systems, with the sole difference lying in their real-space connectivity topology. Yet this change in global connectivity has dramatic spectral consequences. The PBC spectrum of the Möbius ring appears in the complex plane as a multi-petalled rose curve, in sharp contrast to the two disjoint loops of the parallel-ring reference. More crucially, the Möbius reconnection promotes the interchain coupling $t_{mo}$, itself purely reciprocal and topologically inert in isolation, into a control parameter for spectral winding. Once the reconnection is established, increasing $t_{mo}$ drives the spectral winding number through successive quantized topological transitions, with critical coupling values obtainable in closed analytic form, while the winding number of the parallel-ring reference remains topologically rigid. Real-space topological connectivity thus provides an independent route for reshaping spectral winding beyond the non-Hermitian parameters that generate it, and this correspondence holds potential for high-sensitivity sensing [34], non-reciprocal signal amplification, and reconfigurable topological devices.

## II. From Parallel HN Rings to the Möbius Ring System

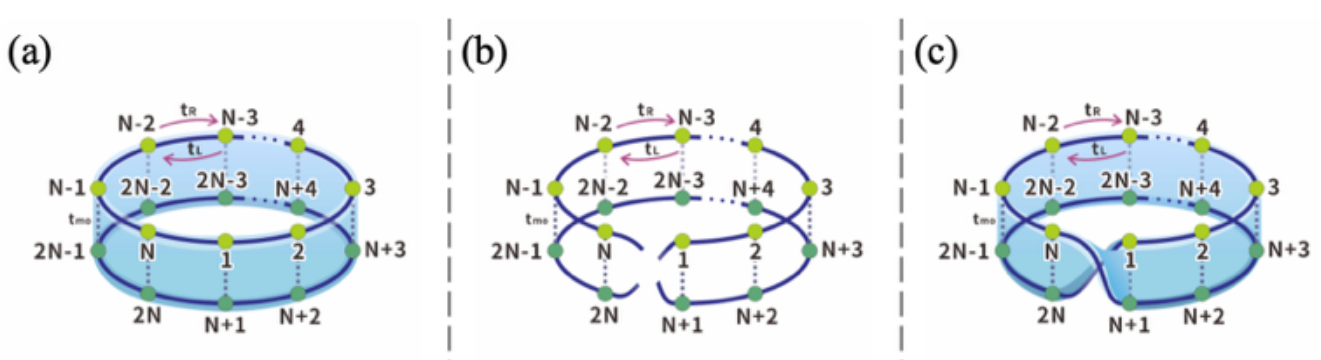

**FIG. 1.** Schematic of the two-parallel-ring system and the Möbius ring system. (a) Two-parallel-ring system ($H_1$): two parallel HN rings with non-reciprocal hopping $t_R, t_L$and reciprocal coupling $t_{mo}$.(b) Topological deformation: the rings are cut at a pair of inter-ring bonds and subjected to a half-twist. (c) Möbius ring system ($H_2$): the two sub-chains are topologically merged into a single closed loop.

As illustrated in Fig. 1a, the two-parallel-ring system consists of two HN [15] rings with asymmetric nearest-neighbor hopping $t_R$ (rightward) / $t_L$ (leftward) and reciprocal inter-ring coupling $t_{mo}$. Its Hamiltonian $H_1$ reads:

$$H_1 = \sum_j v c_j^\dagger c_j + \sum_{\langle i,j\rangle,\text{upper}} \left(t_L c_i^\dagger c_j + t_R c_j^\dagger c_i\right) + \sum_{\langle i,j\rangle,\text{lower}} \left(t_L c_i^\dagger c_j + t_R c_j^\dagger c_i\right) + t_{mo} \sum_j \left(c_j^\dagger c_{j+N} + h.c.\right) \quad (1)$$

where $v$ is the on-site potential, $c_j^\dagger$ $(c_j)$ are fermionic creation (annihilation) operators, $\langle i,j\rangle$ runs over nearest-neighbor pairs within each ring, and $t_{mo}$ couples each site $j$ to its counterpart $j+N$. The non-reciprocity $t_R \neq t_L$ is the hallmark of the HN model.

The central construction (Fig.1 b,c) cuts the two-rings at a pair of inter-ring bonds, reconnects them cross-wise, and closes them into a single Möbius loop, geometrically analogous to joining the ends of a ribbon after a half-twist. This procedure does not alter any local hopping parameter, but fundamentally changes the global boundary condition, generating a non-trivial real-space topology. The resulting Möbius ring system is described by the Hamiltonian $H_2$:

$$H_2 = \sum_j v c_j^\dagger c_j + \sum_{\langle i,j\rangle} (t_L c_i^\dagger c_j + t_R c_j^\dagger c_i) + t_{mo} \sum_j (c_j^\dagger c_{j+N} + h.c.) \quad (2)$$

where indices are taken modulo $2N$, $\langle i,j\rangle$ runs over nearest-neighbor pairs within the big Möbius ring, $t_R$ and $t_L$ are the rightward and leftward hopping amplitudes, and $t_{mo}$ couples each site $j$ to its partner $j+N$.

### III. Möbius Reconnection: Spectral Reshaping and Heterogeneous Skin Localization

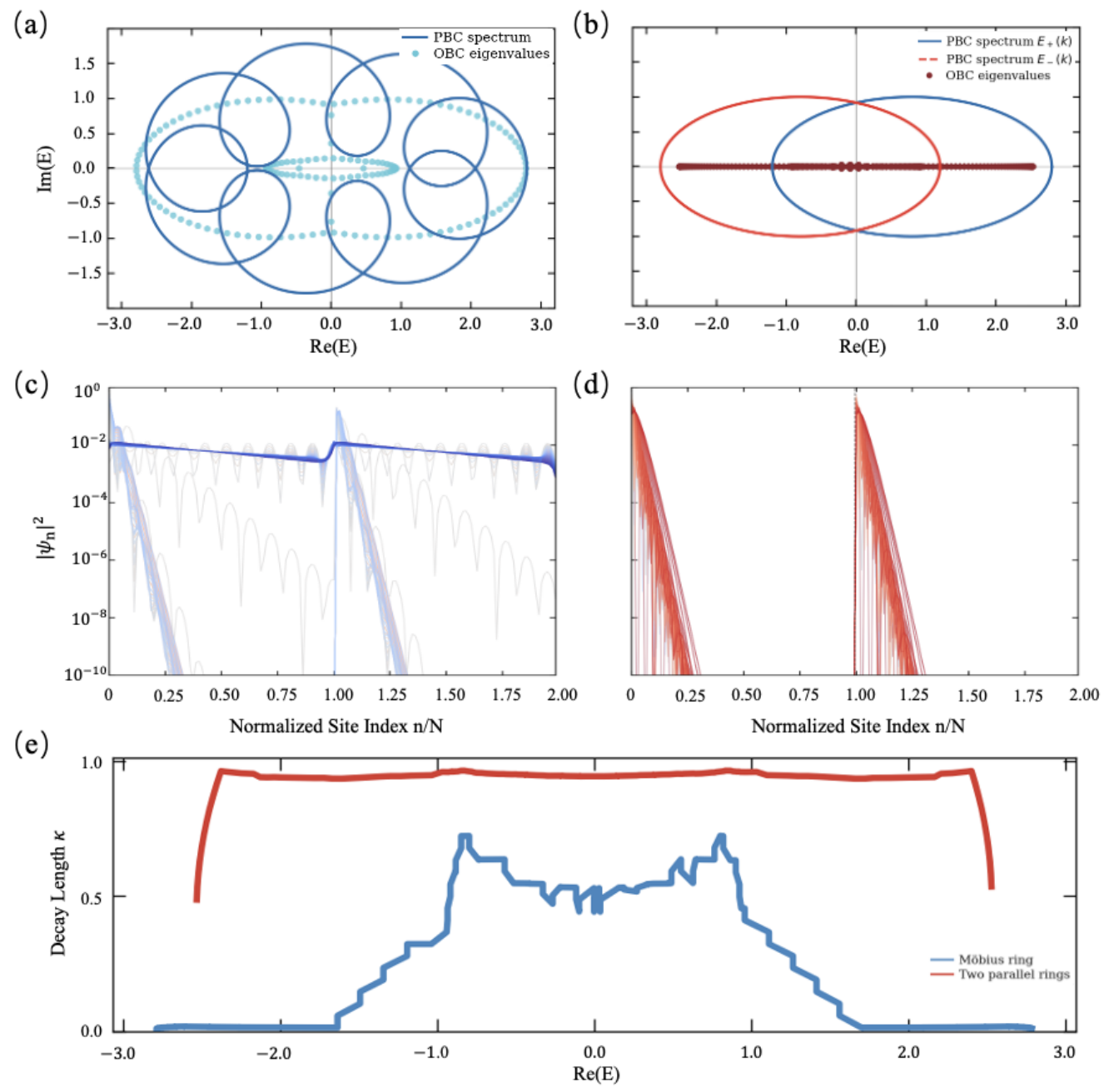

**FIG. 2.** PBC spectrum, OBC skin effect, and decay-length statistics for the Möbius ring system ($H_2$) and the two-parallel-ring system ($H_1$), with $t_L = 1.5$, $t_R = 0.5$, $t_{mo} = 0.8$. (a) Möbius ring system and (b) two-parallel-ring system: PBC continuum (solid curves, $N = 8$ for visual clarity; other values yield the same topology) and OBC eigenvalues (scatter points, $N = 80$ to approximate the thermodynamic limit; larger $N$ gives the same qualitative behavior); the PBC spectrum of the Möbius ring system forms a rose-curve, while that of the two-ring system splits into an ellipse pair $E_\pm(k)$. (c,d) OBC skin-mode intensity profiles $|\psi|^2$ on a logarithmic scale ($N = 80$). Line color encodes the decay length κ; darker lines correspond to larger κ. (e) Decay length κ vs Re(E) for all OBC eigenstates of the Möbius system (blue) and the two-ring system (red) ($N = 80$). Due to the discrete cyclic symmetry, opening the ring at any single bond yields the same OBC spectrum; here we cut the Möbius system between sites 1 and $2N$, and the parallel ring system between sites 1 and $N$ on the first ring and between sites $N+1$ and $2N$ on the second.

The real-space topological difference between the Möbius and two-ring systems is directly imprinted onto their PBC spectra (Fig.2 a,b). Both systems exhibit the skin effect, with their PBC and OBC spectra being fully disjoint, but its nature differs fundamentally. The Möbius system has OBC eigenvalues broadly distributed across the complex plane with widely varying imaginary parts, while those of the parallel ring system are all real, consistent with standard single-chain HN behavior. Notably, the Möbius system retains the discrete cyclic symmetry of the two-ring system, as the OBC eigenvalue spectrum is independent of where the ring is cut, with only the spatial accumulation position of the skin modes shifting.

This contrast is made explicit in the real-space skin-mode intensity profiles (Fig.2 c,d). For the Möbius ring system, skin modes spread across both sub-chains with highly heterogeneous intensity profiles, with different eigenstates decaying at visibly different rates; for the parallel ring system, all eigenstates sharply accumulate at the chain boundary with a nearly uniform spatial envelope.

The energy-resolved decay length $\kappa$ plotted against $Re(E)$ provides the sharpest quantitative distinction [Fig.2 e]. For the Möbius ring system (blue), $\kappa$ is broadly and continuously distributed across a wide interval, varying strongly and non-monotonically with $Re(E)$. This wide spread of $\kappa$ is a defining signature of scale-free localization. Different eigenstates are localized at intrinsically different length scales, and no single decay length characterizes the skin effect. For the two-parallel-ring system (red), most eigenstates share nearly the same $\kappa$ value, characteristic of a universal decay scale.

We can verify that this scale-free behavior is intrinsic rather than a finite-size artifact. As the system size $N$ increases, if Fig.2 c,d were replotted, the intensity envelope of the Möbius ring system would remain unchanged, with the spread in localization lengths persisting across all system sizes; while the envelope of the two-ring system would become progressively steeper, ultimately converging to a single characteristic decay scale. The decay-length distribution shows no dependence on system size, even as the system approaches the thermodynamic limit. This serves as definitive evidence for scale-free localization, in which no characteristic length scale emerges.

### IV. Winding-Number Jumps Driven by Coupling Strength

The connection between the skin effect and the spectral winding number is not merely phenomenological. It has been established rigorously that a nonzero spectral winding number $w(E_0)$ with respect to a reference energy $E_0$ is both necessary and sufficient for the existence of OBC skin modes [9]. This bulk–skin correspondence establishes the winding number as the topological criterion for the skin effect. We now show that in the Möbius ring system, the winding number can be actively tuned by varying the coupling parameters, while the parallel-ring remains topologically rigid under the same operation.

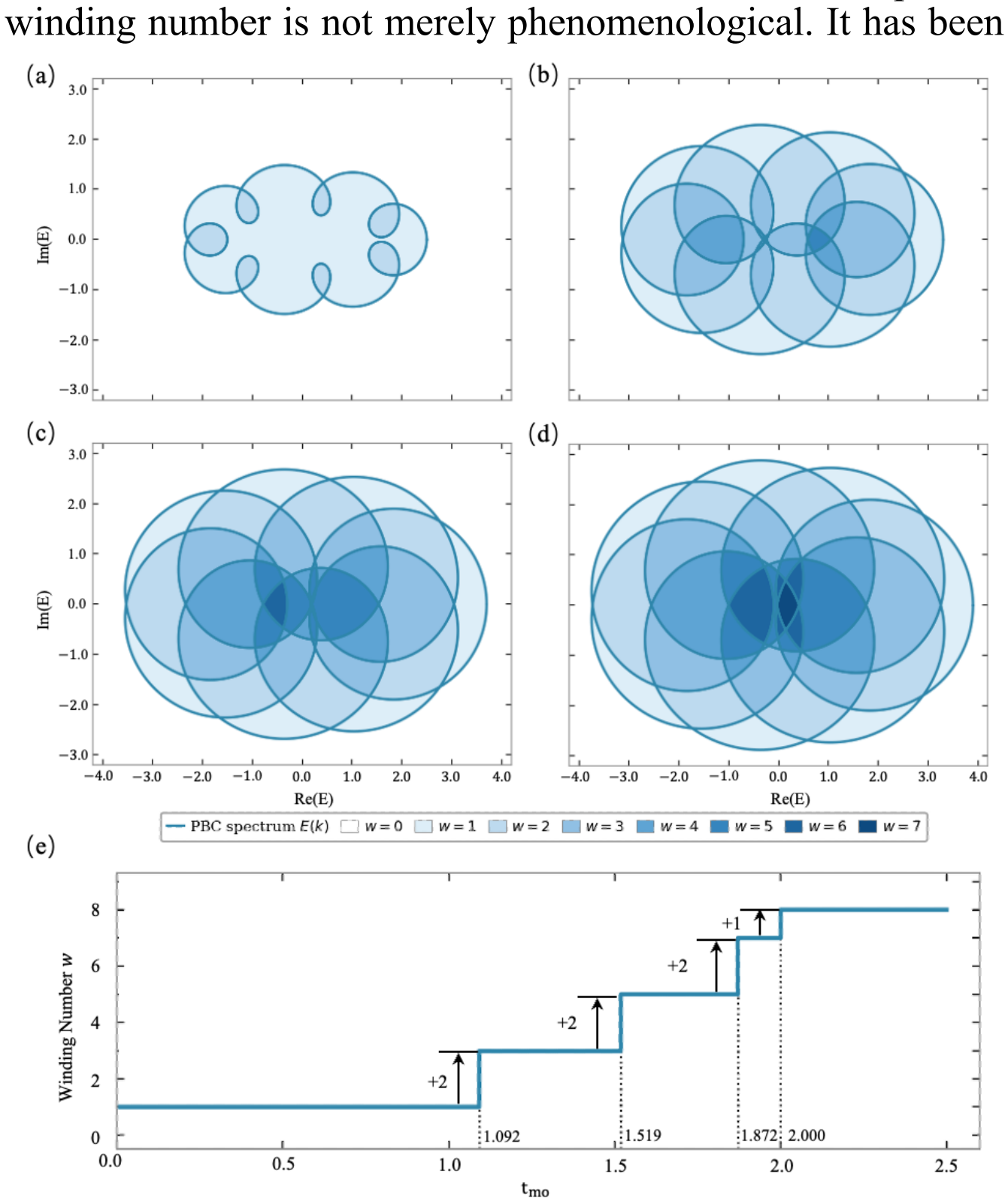

**FIG. 3.** Topological phase transitions in the PBC spectrum of the Möbius ring system as a function of $t_{mo}$, with $t_L = 1.5$, $t_R = 0.5$. (a–d) PBC spectra of the Möbius ring system at representative values of $t_{mo}$; shaded regions indicate the winding number $w$ (see legend). (e) Spectral winding number $w$ vs $t_{mo}$ computed with reference energy $E_0 = 0$; transition points $t_{mo} \approx 1.092, 1.519, 1.872, 2.000$ are marked by dashed lines, with jump magnitudes annotated.

As $t_{mo}$ increases, the PBC spectrum of the Möbius ring system undergoes a series of discrete topological phase transitions (Fig.3 a–d), with the number of rose-curve petals increasing accordingly. Fig.3 e quantifies the winding number computed with reference energy $E_0 = 0$, showing discrete jumps of $+2$ at $t_{mo} \approx 1.092, 1.519, 1.872$, followed by a final jump of $+1$ at $t_{mo} \approx 2.000$. The origin of this behavior lies in the Möbius topology itself. The winding number image of the two-ring system remains unchanged throughout (its shape consistent with the PBC spectrum in Fig.2 b); while upon twisting into the Möbius ring system, the cross-link coupling acquires the ability to continuously reconfigure the global winding structure as $t_{mo}$ varies.

The transition points admit an analytic prediction. The critical momenta $k^*$ satisfy the self-consistent iterative map

$$k^{(j+1)} = \frac{m\pi + \arctan\left(\varepsilon \tan k^{(j)}\right)}{N} \tag{3}$$

initialized at $k_n^{(0)} = m\pi/N$, where $m$ is a positive integer, and $\varepsilon = (t_L - t_R)/(t_L + t_R)$ parametrizes the non-reciprocity. Each value of $m$ selects a distinct branch and converges to a unique $k^*$. For even $N$, the $N/2$ topological transitions correspond to $m = 1, 3, \cdots, N-3$ (yielding $k^* \in (0, \pi/2)$) and $m = N-2$ (yielding $k^* \in (\pi/2, \pi)$), together with the exact terminal solution $k^* = \pi$ (i.e., $m = N$). Substituting the converged $k^*$ into the amplitude condition yields the critical coupling strengths

$$t^*_{mo,m} = \sqrt{(t_L - t_R)^2 + 4 t_L t_R \cos^2(k^*)} \tag{4}$$

which agree with numerical binary-search results to within 1%. For $t_L = 1.5$, $t_R = 0.5$, $N = 8$, the four transitions are recovered at $m$ = 3, 6, 1, 8, giving $t^*_{mo} \approx 1.092$, $1.519$, $1.872$, $2.000$ respectively.

More broadly, in the Möbius ring system, the winding number or the non-Hermitian skin effect can be controlled by tuning the coupling parameters of the system. Among these, $t_{mo}$ is particularly natural to vary; in a circuit or waveguide realization, it can be adjusted by changing the distance between coupled elements, for instance through mutual inductance in an LC resonator array. This establishes a direct and computationally verified mapping from a tunable real-space parameter to a topological invariant, and thence to a measurable localization property—demonstrating that the richness of Möbius non-Hermitian physics is not merely a theoretical curiosity but a controllable physical resource.

## V. Conclusion

In summary, we have proposed and studied the Möbius ring system, revealing fundamental distinctions from the two-parallel-ring system in non-Hermitian physics. The Möbius topology in real space alters the topological structure of the effective Brillouin zone, transforming the PBC spectrum from an ellipse pair composed of two branches $E_\pm(k)$ into a rose-curve, and induces scale-free localization under OBC—different eigenstates are localized at the boundary with intrinsically different decay lengths, a property that persists as the system size increases and stands in sharp contrast to the universal skin behavior of ordinary HN chains. Furthermore, we have shown that the spectral winding number of the Möbius ring system undergoes discrete jumps as the coupling parameter is continuously tuned, steering the system through a series of discrete topological phase transitions. These results establish the Möbius non-Hermitian system as a novel theoretical platform for topologically tunable skin effects. Prospectively, if a physical observable can be identified that changes abruptly at each winding-number jump, it would open the possibility of a new class of non-Hermitian sensing schemes rooted in discrete winding-number jumps, offering threshold sensitivity to parameter variations.